\documentclass[conference]{IEEEtran}
\IEEEoverridecommandlockouts
\usepackage{cite}
\usepackage{amsmath,amssymb,amsfonts}
\usepackage{algorithmic}
\usepackage{graphicx}
\usepackage{textcomp}
\usepackage{xcolor}
\usepackage[caption=false,font=footnotesize]{subfig}
\usepackage{stfloats}
\usepackage{epstopdf}
\usepackage{acronym}
\usepackage[bookmarksopen=true]{hyperref}
\usepackage{mathrsfs}
\usepackage{float}
\usepackage{placeins}

\usepackage{flushend} 

\def\BibTeX{{\rm B\kern-.05em{\sc i\kern-.025em b}\kern-.08em
		T\kern-.1667em\lower.7ex\hbox{E}\kern-.125emX}}

\newcommand{\bs}{\boldsymbol}
\newcommand{\e}{\mathrm{e}}
\newcommand{\C}{\mathbb{C}}
\newcommand{\codeZ}{\mathscr{Z}}
\newcommand{\lp}{\left(}
\newcommand{\rp}{\right)}
\newcommand{\bAlpha}{\bs{\alpha}}
\newcommand{\V}{\mathrm{V}}
\newcommand{\Nidft}{{N}_{\rm idft}}
\newcommand{\PAPR}{\mathrm{PAPR}}
\newcommand{\Pavg}{P_\mathrm{avg}}

\newcommand{\trans}{^\mathrm{T}}
\newcommand{\herm}{^\mathrm{H}}

\DeclareMathOperator*{\argmin}{arg\,min}
\DeclareMathOperator*{\argmax}{arg\,max}
\acrodef{WSN}{wireless sensor network}
\acrodef{USRP}{universal software radio peripheral}
\acrodef{SN}{sensor node}
\acrodef{FC}{fusion center}
\acrodef{MAC}{multiple-access channel}
\acrodef{FL}{federated learning}
\acrodef{ED}{edge device}
\acrodef{CS}{compressed sensing}
\acrodef{ES}[BS]{base station}
\acrodef{DCN}{data center network}
\acrodef{RIS}{reconfigurable intelligent surfaces}
\acrodef{IMC}{in-memory computing}
\acrodef{FPGA}{field-programmable gate array}
\acrodef{SDR}{software-defined radio}
\acrodef{PS}{processing system}
\acrodef{SS}{soft synchronization}
\acrodef{IQ}{in-phase/quadrature}
\acrodef{IP}{intellectual property}
\acrodef{DMA}{direct-memory access}
\acrodef{RAM}{random access memory}
\acrodef{CC}{companion computer}
\acrodef{FEE}{function estimation error}
\acrodef{MSK}{minimum-shift keying}
\acrodef{TDMA}{time-domain multiple access}
\acrodef{PLNC}{physical-layer network coding}
\acrodef{UAV}{unmanned aerial vehicle}
\acrodef{LoRa}{Long-Range}
\acrodef{DC}{direct-current}
\acrodef{DAC}{digital-to-analog converter}
\acrodef{ADC}{anlog-to-digital converter}
\acrodef{CS}{complementary sequence}
\acrodef{GCP}{Golay complementary pair}
\acrodef{ANF}{algebraic normal form}
\acrodef{AACF}{aperiodic auto-correlation function}
\acrodef{RM}{Reed-Muller}
\acrodef{MOCZ}{modulation on conjugate-reciprocal zeros}
\acrodef{BMOCZ}{binary modulation on conjugate-reciprocal zeros}
\acrodef{dizet}[DiZeT]{direct zero-testing}

\acrodef{PUCCH}{physical uplink control channel}
\acrodef{PRACH}{physical random access channel}

\acrodef{OBO}{output-power back-off}
\acrodef{ACLR}{adjacent-channel-leakage ratio}

\acrodef{LDPC}{low-density parity check}

\acrodef{PDF}{probability density function}
\acrodef{CDF}{cumulative distribution function}
\acrodef{CCDF}{complementary cumulative distribution function}

\acrodef{TBMA}{type-based multiple access}

\acrodef{MSFE}{mean-squared function error}
\acrodef{FEE}{function-estimation error}
\acrodef{CER}{computation error rate}
\acrodef{BCER}{block-computation error rate}
\acrodef{CFO}{carrier frequency offset}
\acrodef{TO}{time offset}
\acrodef{PO}{phase offset}
\acrodef{RSSI}{received signal strength  information}

\acrodef{STLC}{space-time line code}
\acrodef{CCI}{co-channel interference}
\acrodef{CSIT}[CSIT]{\ac{CSI} at the transmitter}
\acrodef{CSIR}[CSIR]{\ac{CSI} at the receiver}
\acrodef{MIMO}{multiple-input-multiple-output}
\acrodef{PC}{phase correction}
\acrodef{ZF}{zero-forcing}
\acrodef{ANOVA}{analysis of variance}

\acrodef{PCA}{principal component analysis}
\acrodef{TIG}{Technical Interest Group}

\acrodef{FSK}{frequency-shift keying}
\acrodef{PPM}{pulse-position modulation}
\acrodef{PAM}{pulse-amplitude modulation}

\acrodef{MRC}{maximum-ratio combining}
\acrodef{HP}{hard-coded participation}
\acrodef{HPA}{hard-coded participation with absentees}
\acrodef{SP}{soft-coded participation}
\acrodef{FSK-MV}{\ac{FSK}-based \ac{MV}}
\acrodef{RF}{radio-frequency}
\acrodef{MF}{matched filter}
\acrodef{PPM}{pulse-position modulation}
\acrodef{CSK}{chirp-shift keying}
\acrodef{PPM-MV}[PPM-MV]{\ac{PPM}-based \ac{MV}}
\acrodef{DFT-s-OFDM}{discrete Fourier transform-spread orthogonal frequency division multiplexing}
\acrodef{SC}{single-carrier}
\acrodef{SGD}{stochastic gradient descent}
\acrodef{signSGD}{sign stochastic gradient descent}

\acrodef{SL}{split learning}
\acrodef{SNR}{signal-to-noise ratio}
\acrodef{RMSE}{root-mean-squared error}
\acrodef{OFDM}{orthogonal frequency division multiplexing}
\acrodef{DFT}{discrete Fourier transform}
\acrodef{PSK}{phase-shift keying}
\acrodef{QAM}{quadrature amplitude modulation}
\acrodef{QPSK}{quadrature phase-shift keying}
\acrodef{PMEPR}{peak-to-mean envelope power ratio}
\acrodef{BER}{bit error rate}
\acrodef{SNR}{signal-to-noise ratio}
\acrodef{PSD}{power spectral density}
\acrodef{SE}{spectral efficiency}
\acrodef{CP}{cyclic prefix}
\acrodef{AWGN}{additive white Gaussian noise}
\acrodef{CFR}{channel frequency response}
\acrodef{CIR}{channel impulse response}
\acrodef{MMSE}{minimum mean-squared error}
\acrodef{LMMSE}{linear minimum mean-squared error}
\acrodef{BPSK}{binary phase shift keying}
\acrodef{BPSK}{quadrature phase shift keying}
\acrodef{BLER}{block error rate}
\acrodef{ML}{maximum likelihood}
\acrodef{PHY}{physical layer}
\acrodef{PA}{power amplifier}
\acrodef{IDFT}{inverse discrete Fourier transform}
\acrodef{DoF}{degrees-of-freedom}
\acrodef{IoT}{Internet-of-Things}
\acrodef{FDE}{frequency-domain equalization}
\acrodef{RF}{radio-frequency}
\acrodef{IM}{index modulation}
\acrodef{MF}{matched filter}
\acrodef{PPM}{pulse-position modulation}

\acrodef{MSE}{mean-squared error}
\acrodef{MRT}{maximum-ratio transmission}
\acrodef{ERC}{equal-ratio combining}
\acrodef{BAA}{broadband analog aggregation}
\acrodef{OBDA}{one-bit broadband digital aggregation}
\acrodef{FEEL}{federated edge learning}
\acrodef{FL}{federated learning}
\acrodef{UL}{uplink}
\acrodef{DL}{downlink}
\acrodef{OAC}{over-the-air computation}
\acrodef{TCI}{truncated-channel inversion}
\acrodef{MV}{majority vote}
\acrodef{CNN}{convolution neural network}
\acrodef{ReLU}{rectified-linear unit}
\acrodef{CSI}{channel state information}
\acrodef{PAPR}{peak-to-average power ratio}
\acrodef{SC}{single-carrier}
\acrodef{iid}[IID]{independent and identically distributed}
\acrodef{RMS}{root-mean-square}
\acrodef{4G}{fourth generation}
\acrodef{5G}{Fifth Generation}
\acrodef{6G}{Sixth Generation}
\acrodef{NR}{New Radio}
\acrodef{LTE}{Long-Term Evolution}
\acrodef{OFDMA}{orthogonal frequency division multiple access}
\acrodef{HARQ}{hybrid automatic repeat request}
\acrodef{D2D}{Device-to-Device}
\acrodef{NOMA}{non-orthogonal multiple access}
\acrodef{OMA}{orthogonal multiple access}

\acrodef{IMT}{International Mobile Telecommunications}
\acrodef{ITU}{International Telecommunication Union}

\acrodef{PDP}{power-delay profile}
\acrodef{TBMA}{type-based multiple access}
\acrodef{ISI}{intersymbol interference}
\acrodef{MLSE}{maximum likelihood sequence estimator}
\acrodef{LTI}{linear time-invariant}

\begin{document}
	
	\title{On the Optimal Radius and Subcarrier Mapping for Binary Modulation on Conjugate-Reciprocal Zeros}
	

	\author{\IEEEauthorblockN{Parker Huggins}
	\IEEEauthorblockA{\textit{Department of Electrical Engineering} \\
		University of South Carolina \\ 
		Columbia, SC, USA \\
		parkerkh@email.sc.edu}
	\and
	\IEEEauthorblockN{Alphan \c{S}ahin}
	\IEEEauthorblockA{\textit{Department of Electrical Engineering} \\
		University of South Carolina \\ 
		Columbia, SC, USA \\
		asahin@mailbox.sc.edu}
	}
	
	\maketitle
	
	\begin{abstract}  	
		In this work, we investigate the radius maximizing reliability for \ac{BMOCZ} implemented with both \ac{ML} and \ac{dizet} decoders. We first show that the optimal radius for \ac{BMOCZ} is a function of the employed decoder and that the radius maximizing the minimum distance between polynomial zeros does not maximize the minimum distance of the final code. While maximizing zero separation offers an almost optimal solution with the \ac{dizet} decoder, simulations show that the \ac{ML} decoder outperforms the \ac{dizet} decoder in both \ac{AWGN} and fading channels when the radius is chosen to maximize codeword separation. Finally, we analyze different sequence-to-subcarrier mappings for \ac{BMOCZ}-based \ac{OFDM}. We highlight a flexible time-frequency \ac{OFDM} waveform that avoids distortion introduced by a frequency-selective channel at the expense of a higher \ac{PAPR}.
	\end{abstract}
	
	\begin{IEEEkeywords}
		Huffman sequences, BMOCZ, polynomial zeros, OFDM, subcarrier mappings, waveforms.
	\end{IEEEkeywords}
	
	\acresetall
	\section{Introduction} \label{sec:intro}
	
	The emergence of the \ac{IoT} has presented an abundance of challenges to modern communication systems. Unlike previous generations of wireless networks focused on enhancing data rate for human users, \ac{5G} \ac{NR} has called for the development of novel use cases, including machine-type \cite{bockelmann2016massive} and ultra-reliable low-latency communication \cite{popovski2014ultra}. Such new demands necessitate flexible communication technologies that are adaptable to different environments, enabling the connection of users, sensors, and devices on an unprecedented scale. Looking towards sixth-generation wireless networks, non-coherent communication strategies have gained traction for their ability to scale with the anticipated stark increase in wireless connectivity \cite{nawaz2021non}. Furthermore, the low complexity and low power consumption of non-coherent based communication hardware lend themselves to application in the \ac{IoT} \cite{witrisal2009noncoherent}, where the billions of devices connected worldwide will pose challenges to current communication infrastructure. A non-coherent communication system is one in which the receiver has no explicit knowledge of \ac{CSI}. In this case, equalization techniques that are commonplace in modern wireless communication systems cannot be used for symbol detection; rather, the receiver must perform ``blind" demodulation of the received signal. The design of non-coherent communication systems that are both reliable and practical thus proves challenging \cite{xu2019sixty}. 
	
	A recently proposed non-coherent communication scheme for short packets is \ac{MOCZ}. Using \ac{MOCZ}, information bits are encoded into the zeros of the baseband signal's $z$-transform. The baseband signal thus takes the form of a polynomial in the $z$-domain, and the transmitted sequence comprises the coefficients of this polynomial \cite{walk2017short}. An advantage of MOCZ is that the zeros of the transmitted polynomial are unaffected by the \ac{CIR}: the convolution of the polynomial coefficients with the \ac{CIR} corresponds to multiplication in the $z$-domain, an operation that may introduce extraneous zeros to the received polynomial but not alter those already transmitted \cite{walk2019principles}. Both the theoretical and practical aspects of \ac{MOCZ} are studied extensively in \cite{walk2019principles} and \cite{walk2020practical}. In particular, a variant of \ac{MOCZ} is introduced whereby each information bit is encoded into the zero of a conjugate-reciprocal zero pair. The technique is termed \ac{BMOCZ}, and it yields polynomial coefficients that are Huffman sequences \cite{ackroyd1970design}. Furthermore, various improvements and applications of \ac{MOCZ} have been considered in the literature. For example, the authors in \cite{siddiqui2023spectrally} propose spectrally-efficient \ac{BMOCZ} using faster-than-Nyquist signaling. In \cite{sasidharan2024alternative}, codebooks are introduced for \ac{MOCZ} that reduce \ac{PAPR}. In \cite{sun2023noncoherent} and \cite{walk2021multi}, the authors investigate diversity techniques and multi-user access for \ac{MOCZ}, respectively.
	
	A crucial parameter in the design of \ac{BMOCZ} is the radius deciding the placement of conjugate-reciprocal zero pairs. The choice of this radius determines the separation between both zero-vectors and codewords, i.e., factors influencing the performance of the communication scheme. For this reason, the radius maximizing zero separation for \ac{BMOCZ} was determined in \cite{walk2019principles} and introduced in conjunction with a \ac{dizet} decoder. To our knowledge, however, the general optimal radius for \ac{BMOCZ} is not addressed. Therefore, in this preliminary work, we investigate the optimal radius for \ac{BMOCZ} implemented using both \ac{ML} and \ac{dizet} decoders. We show that the \ac{ML} decoder outperforms the \ac{dizet} decoder in both \ac{AWGN} and fading channels for appropriate choices of the radius parameter. Moreover, the merits of different subcarrier mapping strategies for \ac{BMOCZ}-based \ac{OFDM} are discussed. Simulations of the proposed \ac{OFDM} waveforms in a fading channel demonstrate that a time-mapping approach achieves the best \ac{BLER} performance at the expense of increased \ac{PAPR}.
	
	\emph{Notation}: The set of complex numbers is denoted by $\C$, and the complex-conjugate of a complex number $z=a+jb$ is expressed as $z^\ast=a-jb$. We denote the Euclidean norm of a vector $\bs{\mathrm{v}}\in\C^{N\times1}$ as $\lVert \bs{\mathrm{v}} \rVert_2=\sqrt{\bs{\mathrm{v}}\herm\bs{\mathrm{v}}}$. The probability of an event $A$ given event $B$ is denoted by $\mathrm{Pr}(A|B)$. The circularly symmetric complex normal distribution with zero-mean and variance $\sigma^2$ is expressed as $\mathcal{CN}(0,\sigma^2)$. The expected value of a random variable $X$ is denoted by $\mathbb{E}[X]$.
	
	\section{Preliminaries} \label{sec:preliminaries}
	
	This section reviews \ac{BMOCZ} and describes the two decoders considered in this work. To begin, consider a binary message $\bs{\mathrm{m}}=(m_0,m_1,\hdots,m_{K-1})$. Using \ac{BMOCZ}, the $K$ message bits are modulated onto $K$ distinct zeros 
	\begin{equation} \label{eq:zeroMapping}
		\alpha_k=
		\begin{cases}
			R \:\e^{j2\pi \frac{k}{K}}, & m_k=1\\
			\frac{1}{R}\:\e^{j2\pi \frac{k}{K}}, & m_k=0\\
		\end{cases},
	\end{equation}
	where $k=0,1,\hdots,K-1$ and $R>1$ is the radius determining the distance between conjugate-reciprocal zero pairs. By the fundamental theorem of algebra, the $K$ zeros define a polynomial of degree $K$, namely,
	\begin{equation} \label{eq:Xz}
		X(z)=x_K\prod_{k=0}^{K-1}(z-\alpha_k),
	\end{equation} 
	where $z\in\C$ and $x_K\neq 0$  is a scalar multiple that does not affect the zero locations of $X(z)$. In discrete-time, the baseband sequence to transmit comprises the $K+1$ polynomial coefficients of $X(z)$, i.e., $\bs{\mathrm{x}}=(x_0,x_1,\hdots,x_K)\trans$. It is typical to choose $x_K$ such that transmitted sequences are normalized with unit energy or, equivalently, such that $\lVert\bs{\mathrm{x}}\rVert_2^2=1$.
	
	For the duration of transmission, it is assumed that the channel is \ac{LTI} with an $L$-tap impulse response $\bs{\mathrm{h}}=(h_0,h_1,\hdots,h_{L-1})\trans$. Using the convolution theorem, the received sequence $\bs{\mathrm{y}}=(y_0,y_1,\hdots,y_{K+L-1})\trans$ can be expressed in the $z$-domain as
	\begin{equation} \label{eq:Yz}
		Y(z)=X(z)H(z)+W(z),
	\end{equation}
	where $H(z)$ and $W(z)$ represent the unilateral $z$-transform of $\bs{\mathrm{h}}$ and a noise sequence $\bs{\mathrm{w}}=(w_0,w_1,\hdots,w_{K+L-1})\trans$, respectively. Note that $H(z)$ and $W(z)$ are both polynomials in the complex variable $z$ and thus can be factored in terms of their zeros. For $N=K+L$, we can write
	\begin{equation} \label{eq:Hz}
		H(z)=h_{L-1}\prod_{l=0}^{L-2}(z-\beta_l)
	\end{equation} 
	and 
	\begin{equation} \label{eq:Wz}
		W(z)=w_{N-1}\prod_{n=0}^{N-2}(z-\gamma_n).
	\end{equation} 
	Therefore, the polynomial in (\ref{eq:Yz}) has a total of $N-1$ zeros, $K$ of which are information-bearing.
	
	The authors in \cite{walk2019principles} propose several methods for demodulating and decoding the received polynomial sequence. In particular, introduced are an \ac{ML} and \ac{dizet}  decoder. The \ac{ML} decoder estimates the transmitted zeros by searching over all possible zero-vectors $\bAlpha\in\codeZ^K$, where $\codeZ^K$ is the \ac{BMOCZ} zero-codebook for a given $K$; the codebook is generated by taking the Cartesian product of all conjugate-reciprocal zero pairs $\codeZ_k=\{\alpha_k,1/\alpha_k^\ast\}$, i.e., $\codeZ^K=\codeZ_0\times\codeZ_1\times\cdots\times\codeZ_{K-1}$. Using the \ac{ML} decoder, assuming a uniform \ac{PDP} \cite{walk2019principles}, an estimate of the transmitted zeros is obtained directly as
	\begin{equation} \label{eq:ML}
		\hat{\bAlpha}=\argmin_{\bAlpha\in\codeZ^K}\left\lVert \lp \bs{\V}_{\bAlpha}\herm \bs{\V}_{\bAlpha} \rp^{-\frac{1}{2}}\bs{\V}_{\bAlpha}\herm\bs{\mathrm{y}} \right\rVert_2^2,
	\end{equation}
	where $\bs{\V}_{\bAlpha}\herm$ is the $K\times N$ Vandermonde matrix 
	\begin{equation} \label{eq:vandermonde}
		\bs{\V}_{\bAlpha}\herm=
		\begin{pmatrix}
			1 & \alpha_1 & \alpha_1^2 & \hdots & \alpha_1^{N-1}\\
			1 & \alpha_2 & \alpha_2^2 & \hdots & \alpha_2^{N-1}\\
			\vdots & \vdots & \vdots & \ddots & \vdots\\
			1 & \alpha_K & \alpha_K^2 & \hdots & \alpha_K^{N-1}
		\end{pmatrix}.
	\end{equation}
	Instead of searching across all $\bAlpha\in\codeZ^K$, the \ac{dizet} decoder simply evaluates the received polynomial in (\ref{eq:Yz}) at the zeros in $\codeZ_k$. The $k$th transmitted zero is then estimated as 
	\begin{equation} \label{eq:DiZeT}
		\hat{\alpha}_k=\argmin_{\alpha\in\codeZ_k}\left| Y(\alpha) \right|\cdot |\alpha|^{-\frac{N-1}{2}},
	\end{equation}
	where the weighting factor $|\alpha|^{-(N-1)/2}$ is introduced to scale the output of $|Y(\alpha)|$ to balance the exponential nature of the polynomial coefficients \cite{walk2019principles}.
	
	\section{Design Considerations} \label{sec:optimal_radius}
	
	\subsection{Optimal Radius} \label{subsec:radius}
	
	The form of (\ref{eq:zeroMapping}) raises a natural question: for a given $K$, what is the radius $R$ that maximizes the reliability of \ac{BMOCZ}? The accepted answer in current literature is the radius which maximizes the separation between zeros  \cite{walk2019principles,walk2020practical,siddiqui2023spectrally}, i.e., 
	\begin{equation} \label{eq:dizetR}
		R_\mathrm{DZ}(K)=\sqrt{1+\sin(\pi/K)}.
	\end{equation}
	This result is intuitive for \ac{BMOCZ} employing the \ac{dizet} decoder, for the received polynomial sequence is directly evaluated at the possible zero locations. The \ac{ML} decoder, however, does not involve the explicit evaluation of (\ref{eq:Yz}) at any zeros. Moreover, the \ac{ML} decoder is derived from the general \ac{MLSE} \cite{forney1972maximum} given by 
	\begin{equation} \label{eq:mlse}
		\hat{\bs{\mathrm{x}}}=\argmax_{\bs{\mathrm{x}}\in\mathscr{C}^K}\mathrm{Pr}(\bs{\mathrm{y}}|\bs{\mathrm{x}}),
	\end{equation} 
	where $\mathscr{C}^K$ is the \ac{BMOCZ} polynomial-codebook for a given $K$ \cite{walk2019principles}. Since the optimization is performed over polynomial sequences and not zeros, it is best to choose the radius that maximizes the separation between codewords, i.e.,
	\begin{equation} \label{eq:mlR}
		R_\mathrm{ML}(K)=\argmax_{R>1}\lp\min_{\bs{\mathrm{x}}_i,\bs{\mathrm{x}}_j\in\mathscr{C}^K}\left\lVert\bs{\mathrm{x}}_i-\bs{\mathrm{x}}_j\right\rVert_2^2\rp, \ i\neq j.
	\end{equation}
	
\begin{figure}[t]
		\centering
		\subfloat[Minimum codeword separation against $R$ values.]{\includegraphics[width=3in]{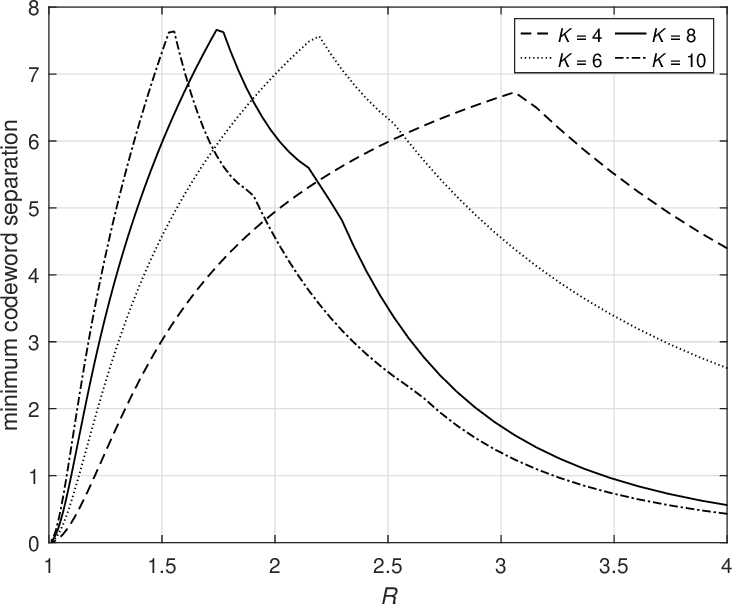}\label{subfig:RvsDmin}}~\\
		\subfloat[Optimal $R$ for various $K$ values.]{\includegraphics[width=3in]{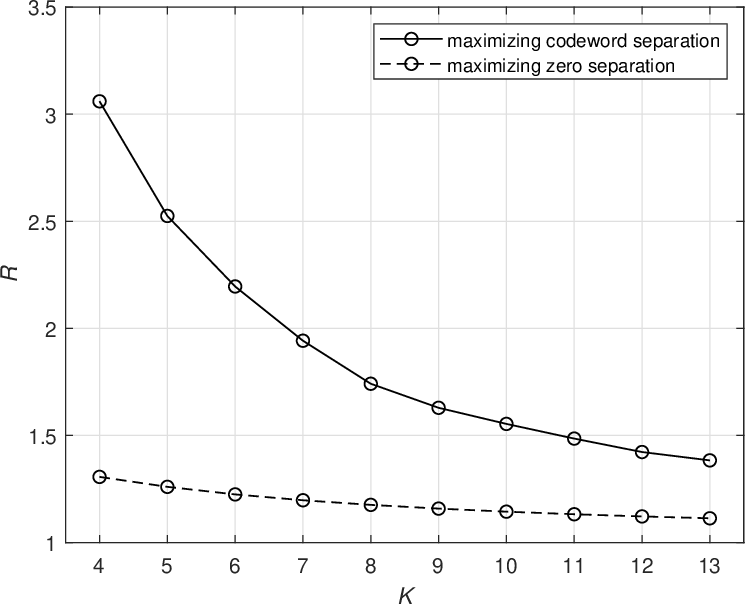}\label{subfig:KvsR}}	
		\caption{Minimum codeword separation and the radii maximizing zero and codeword separation for select $K$.}	
		\label{fig:KandR}
	\end{figure}

	To compare the radii given in (\ref{eq:dizetR}) and (\ref{eq:mlR}), we generate \ac{BMOCZ} zero- and polynomial-codebooks for various $K$. Fig.~\ref{fig:KandR}(a) displays the minimum observed codeword separation for $K\in\{4,6,8,10\}$ and $R\in(1,4]$. Fig.~\ref{fig:KandR}(b) shows the corresponding radii maximizing zero and codeword separation for $K\in\{4,5,\hdots,13\}$, i.e., the radii computed in accordance with (\ref{eq:dizetR}) and (\ref{eq:mlR}), respectively. For all $K$, notice that the radius maximizing codeword separation is greater than that maximizing zero separation. It follows that the optimal radius for \ac{BMOCZ} is a function of the utilized decoder.
	
	\subsection{OFDM Subcarrier Mappings} \label{subsec:mappings}
	
	\def\dataSymbols[#1]{d_{#1}}
	\def\subcarrierIndex{\ell}
	\def\polynomialIndex{p}
	\def\ofdmSymbolIndex{m}
	\def\sampleIndex{n}
	\def\sequenceIndex{k}
	
	\def\numberOfSubcarriers{\mathcal{L}}
	\def\cpSize{N_{\rm cp}}
	\def\numberOfPolynomials{P}
	\def\ofdmSample[#1][#2]{s_{#1}[#2]}
	
	\begin{figure}[t]
		\centering
		\subfloat[Time-mapping.]{\includegraphics[width=2in]{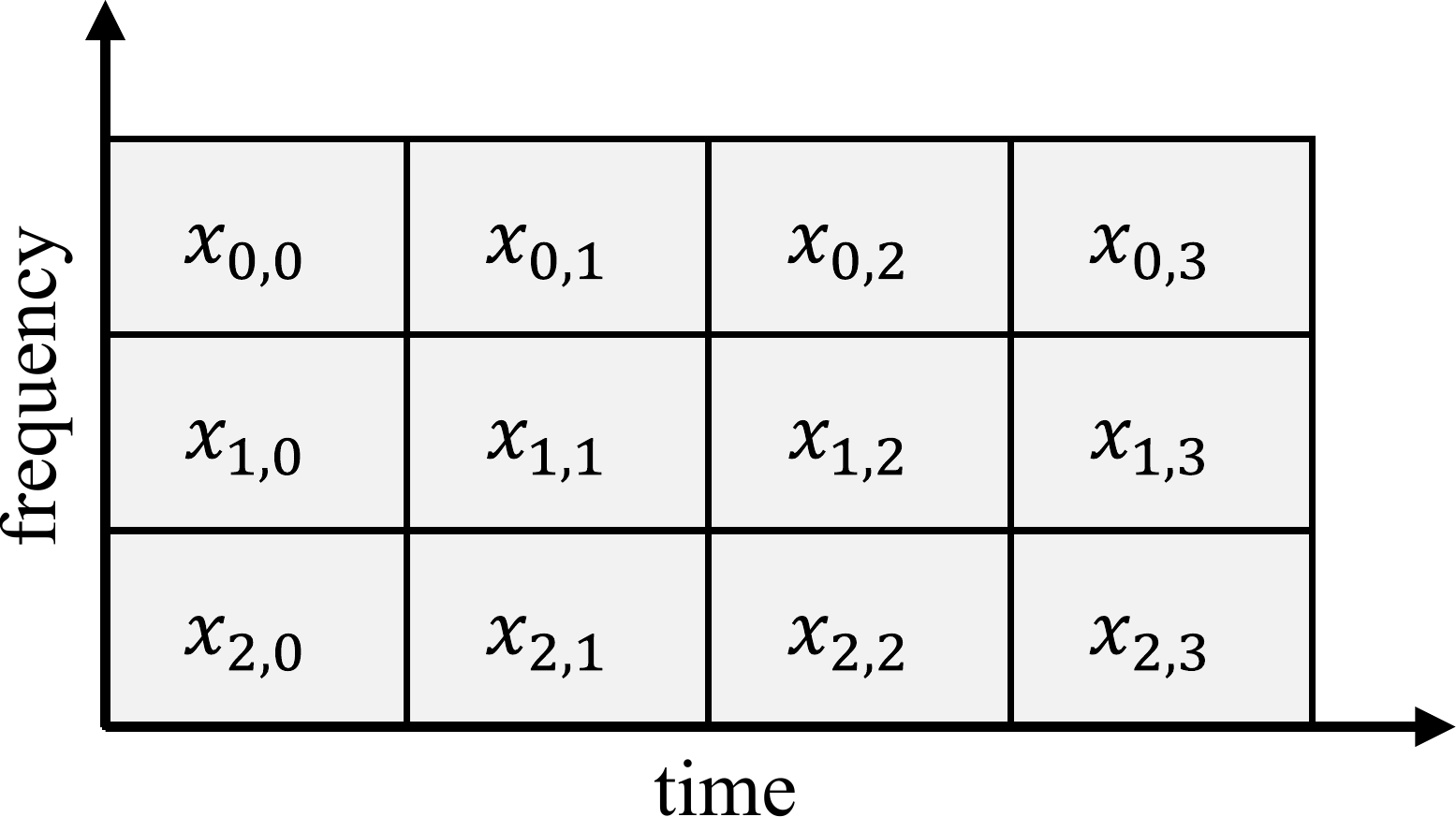}\label{subfig:timeMapping}}~\\
		\subfloat[Frequency-mapping.]{\includegraphics[width=1.5in]{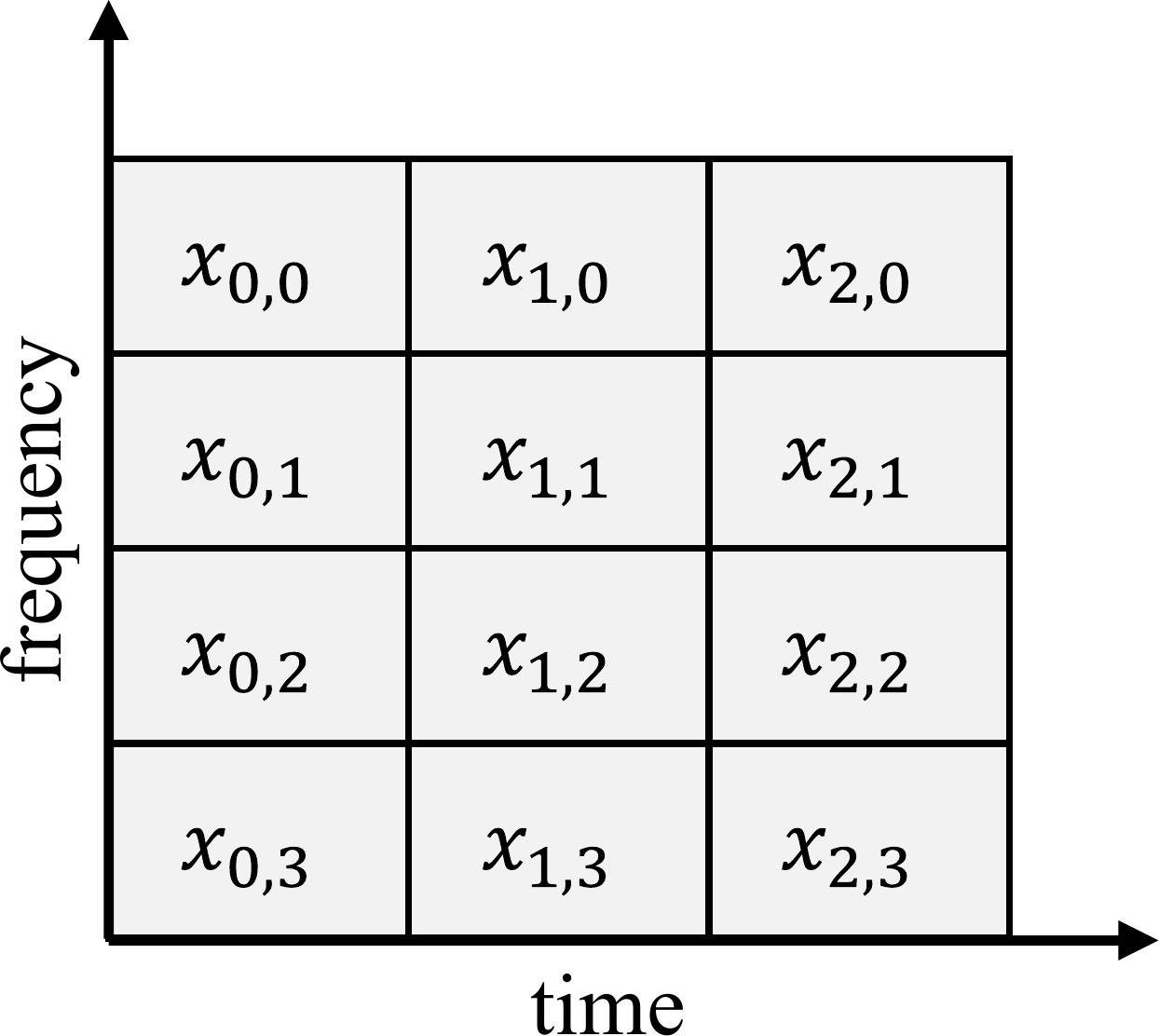}\label{subfig:freqMapping}}~
		\hspace{0.5cm}
		\subfloat[Time-frequency mapping.]{\includegraphics[width=1.5in]{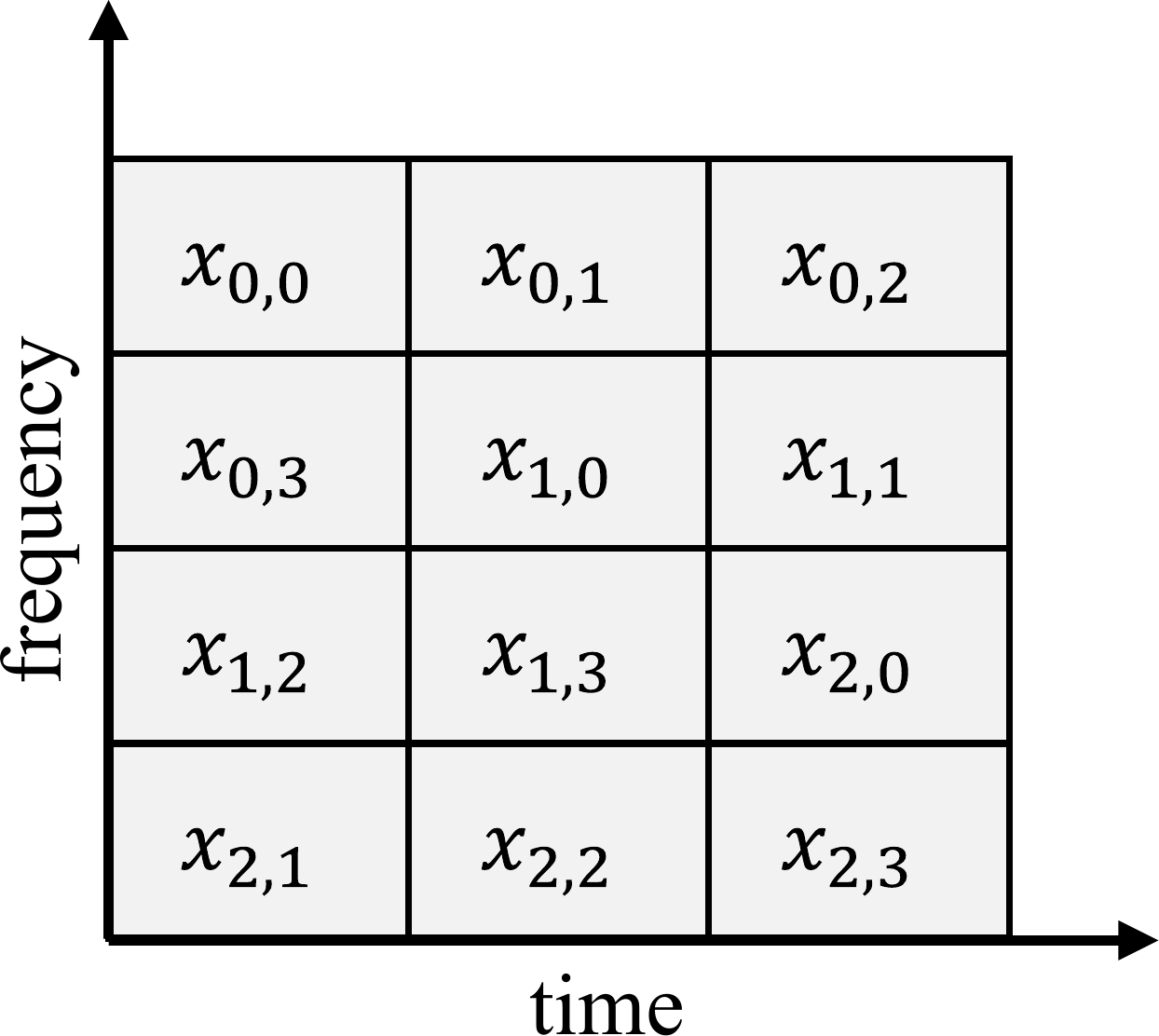}\label{subfig:timeFreqMapping}}
		\caption{Examples of the three principal sequence-to-subcarrier mappings for \ac{BMOCZ}-based \ac{OFDM} with $P=3$, $K=3$, and $M=3$.}
		\label{fig:ofdmMappings}
	\end{figure}
	
	This section outlines how we construct \ac{BMOCZ}-based \ac{OFDM} waveforms to limit the number of additional zeros in (\ref{eq:Yz}) introduced by the channel \cite{walk2021multi}. We begin by considering a binary message of length $B\geq K$. Next, $\numberOfPolynomials$ polynomials are generated by mapping every $K$ bits of the message to sequences via \ac{BMOCZ}. Let $x_{\polynomialIndex,\sequenceIndex}$ denote the $k$th coefficient of the $\polynomialIndex$th polynomial for $\polynomialIndex\in\{0,1,\dots,\numberOfPolynomials-1\}$ and  $\sequenceIndex\in\{0,1,\dots,K\}$. In principle, there exists three distinct sequence-to-subcarrier mappings: time-mapping, frequency-mapping, and time-frequency mapping. In what follows, we discuss the merits of each approach.
	
	With time-mapping, $\numberOfPolynomials$ subcarriers are allocated for the polynomial sequences. The $\subcarrierIndex$th subcarrier of $K+1$ \ac{OFDM} symbols then constitutes one of the $\numberOfPolynomials$ polynomials. Specifically, the $\ofdmSymbolIndex$th \ac{OFDM} symbol can be expressed as
	\begin{equation} \label{eq:timeOFDM}
		\ofdmSample[\ofdmSymbolIndex][\sampleIndex]=\frac{1}{\sqrt{\Nidft}}\sum_{\subcarrierIndex=0}^{\numberOfSubcarriers-1}\dataSymbols[\subcarrierIndex,\ofdmSymbolIndex]\e^{j2\pi \subcarrierIndex\frac{\sampleIndex}{\Nidft}}, \ -\cpSize\le \sampleIndex <\Nidft,	
	\end{equation}	
	where $\dataSymbols[\subcarrierIndex,\ofdmSymbolIndex]$ is the data symbol transmitted on the $\subcarrierIndex$th subcarrier of the $\ofdmSymbolIndex$th OFDM symbol, $\Nidft$ is the \ac{IDFT} size, and $\cpSize$ is the \ac{CP} size. For time-mapping,  $\dataSymbols[\subcarrierIndex,\ofdmSymbolIndex]$ is simply the $\sequenceIndex$th polynomial coefficient of the $\polynomialIndex$th polynomial sequence, i.e., $\dataSymbols[\subcarrierIndex,\ofdmSymbolIndex]=x_{\polynomialIndex,\sequenceIndex}$ for $\subcarrierIndex=\polynomialIndex\in\{0,1,\dots,\numberOfPolynomials-1\}$ and $\ofdmSymbolIndex=\sequenceIndex\in\{0,1,\dots,K\}$. Fig.~\ref{fig:ofdmMappings}(a) illustrates the time-mapping approach in the simple case where $P=K=3$. The advantage of time-mapping is that it can accommodate large $K$, provided that the duration of each sequence is less than the coherence time of the channel. Moreover, even in a frequency-selective channel, each polynomial sequence experiences multiplication by a single complex gain due to the \ac{CP}. Hence, no additional zeros are introduced to the transmitted sequences (i.e., $L_\mathrm{eff}=1$).
	
\begin{figure*}[t]
		\centering
		\includegraphics[width=6.5in]{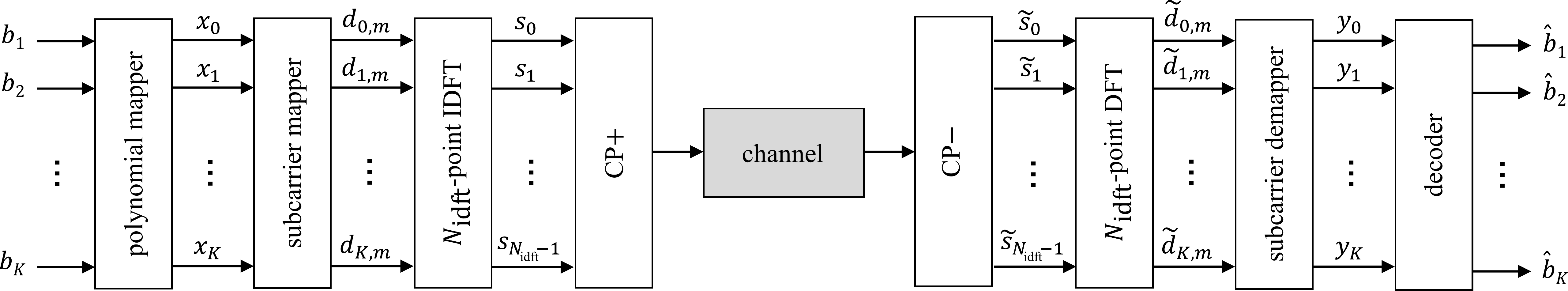}
		\caption{Block diagram showing the processing of the $m$th \ac{OFDM} symbol for a frequency-mapping approach ($m=p$).}	
		\label{fig:ofdmDiagram}
	\end{figure*}

	With frequency-mapping, the polynomial coefficients are mapped to subcarriers directly such that each \ac{OFDM} symbol holds a single polynomial sequence, i.e., $\numberOfSubcarriers=K+1$ and  $\dataSymbols[\subcarrierIndex,\ofdmSymbolIndex]=x_{\polynomialIndex,\sequenceIndex}$ for $\subcarrierIndex=\sequenceIndex$ and $\ofdmSymbolIndex=\polynomialIndex$. A simple example of the frequency-mapping approach is provided in Fig.~\ref{fig:ofdmMappings}(b) for $P=K=3$. Fig.~\ref{fig:ofdmDiagram} depicts the block diagram of a general \ac{BMOCZ}-based \ac{OFDM} scheme implemented with frequency-mapping. Because Huffman sequences have almost ideal aperiodic auto-correlation functions \cite{ackroyd1970design}, mapping the sequences to frequency and not time is beneficial for \ac{PAPR}. However, the disadvantage of this approach is that coefficients of the same polynomial sequence can experience multiplication by different complex gains. The length of transmittable sequences is therefore limited by the coherence bandwidth of the channel, an imposition often much stricter than the coherence time in practice, particularly for low-mobility environments.
	
	Time-frequency mapping utilizes $M$ \ac{OFDM} symbols and $\numberOfSubcarriers=\left\lceil P(K+1)/M\right\rceil$ subcarriers. The data symbol transmitted on the $\subcarrierIndex$th subcarrier of the $\ofdmSymbolIndex$th \ac{OFDM} symbol is $\dataSymbols[\subcarrierIndex,\ofdmSymbolIndex]=x_{\polynomialIndex,\sequenceIndex}$ for $\polynomialIndex=\left\lfloor\frac{\subcarrierIndex M+\ofdmSymbolIndex}{K+1}\right\rfloor$, $k=\mathrm{mod}(\subcarrierIndex M+\ofdmSymbolIndex,K+1)$, $\subcarrierIndex\in\{0,1,\dots,\numberOfSubcarriers-1\}$, and $\ofdmSymbolIndex\in\{0,1,\dots,M-1\}$. An example of time-frequency mapping is depicted in Fig.~\ref{fig:ofdmMappings}(c) for $M=3$ and $\numberOfSubcarriers=4$. The advantage of time-frequency mapping is its flexibility, for either the number of \ac{OFDM} symbols or subcarriers can be freely selected. This is not the case for time- and freqency-mapping approaches, where the number of \ac{OFDM} symbols and the number of subcarriers are fixed to $K+1$, respectively.
	
	\section{Numerical Results} \label{sec:results}
	
	\subsection{Decoder Performance at Optimal Radius} \label{subsec:decoder_performance}
	
	\begin{figure}[tp]
		\centering
		\subfloat[Bit error rate performance  in AWGN.]{\includegraphics[width=3in]{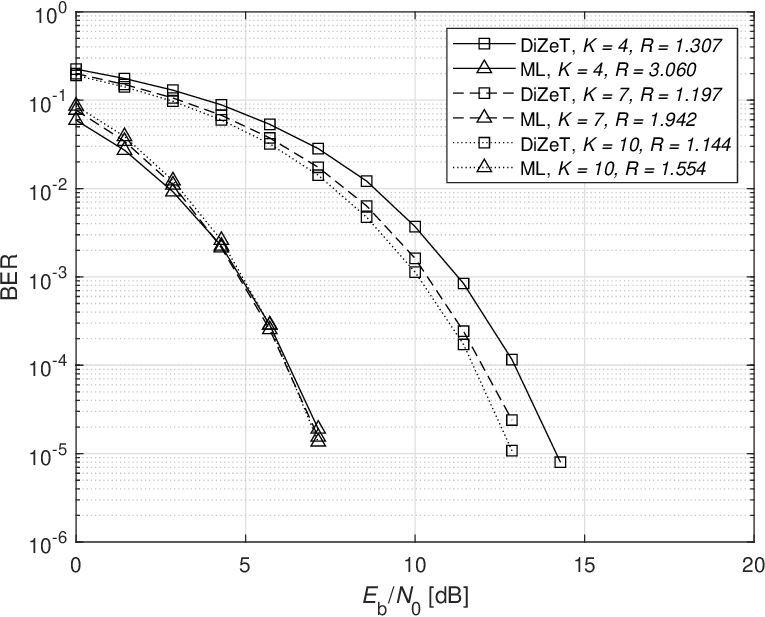}\label{subfig:awgnBER}}~\\
		\subfloat[Block error rate performance in AWGN.]{\includegraphics[width=3in]{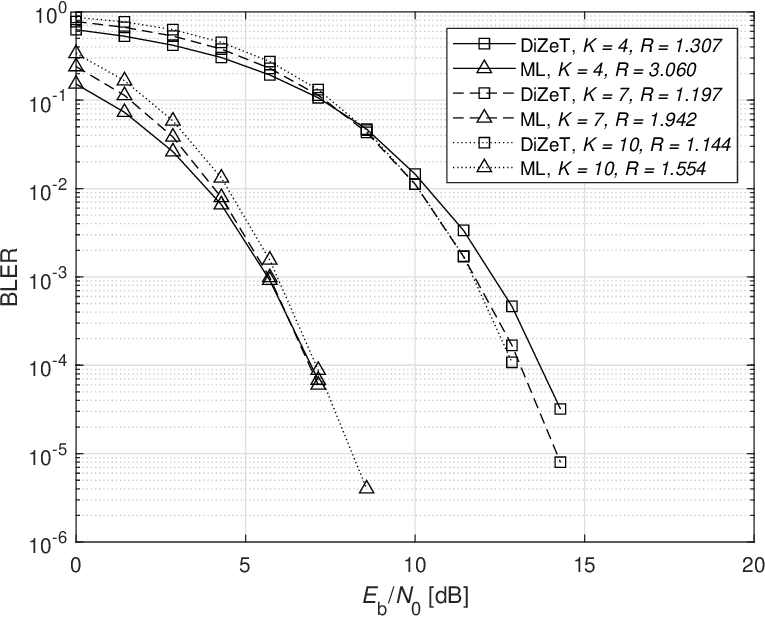}\label{subfig:awgnBLER}}	
		\caption{Comparison of the \ac{ML} and \ac{dizet} decoders in an \ac{AWGN} channel. For each $K$, the radius $R$ is chosen to maximize codeword and zero separation for the \ac{ML} and \ac{dizet} decoders, respectively.}	
		\label{fig:awgnErrorCurves}
	\end{figure}
	
	\begin{figure}[t]
		\centering
		\subfloat[Bit error rate performance in a fading channel.]{\includegraphics[width=3in]{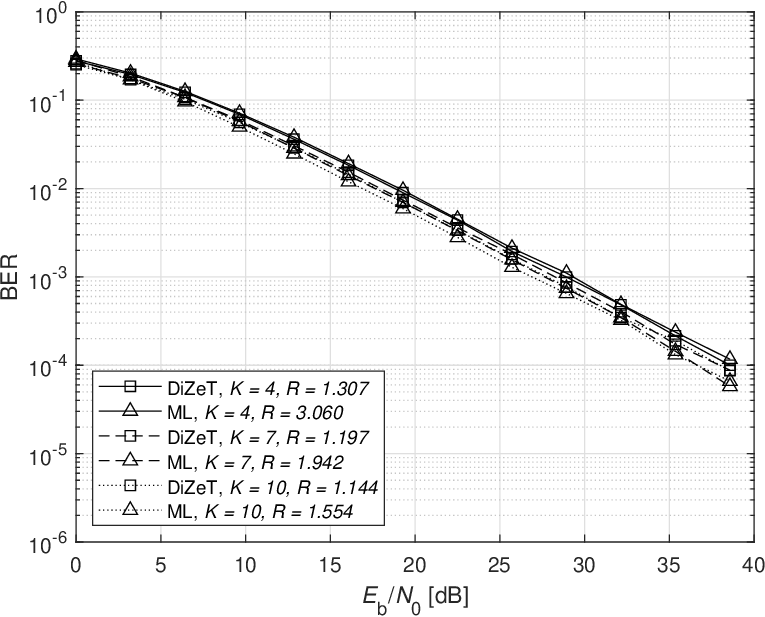}\label{subfig:fadingBER}}~\\
		\subfloat[Block error rate performance in a fading channel.]{\includegraphics[width=3in]{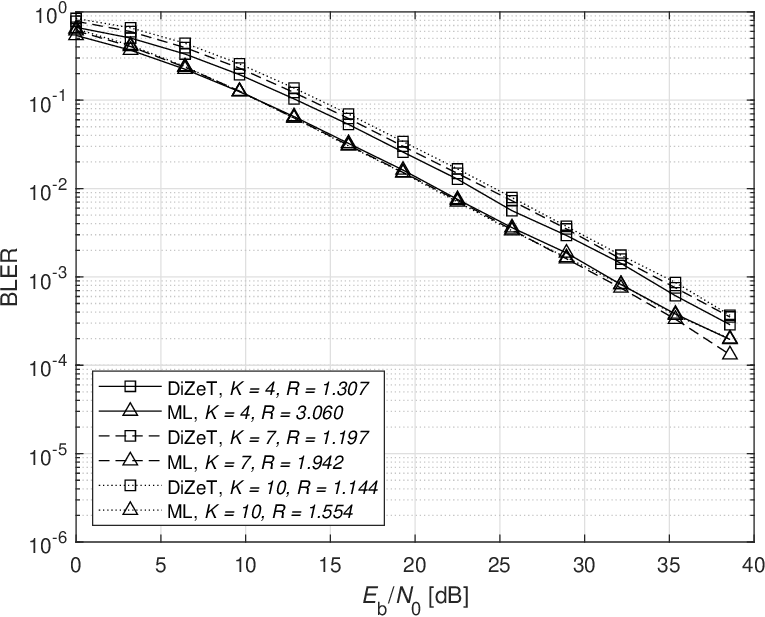}\label{subfig:fadingBLER}}	
		\caption{Comparison of the \ac{ML} and \ac{dizet} decoders in a fading channel with $L=1$. For each $K$, the radius $R$ is chosen to maximize codeword and zero separation for the \ac{ML} and \ac{dizet} decoders, respectively.}	
		\label{fig:fadingErrorCurves}
	\end{figure}
	
	This section compares the performance of a \ac{BMOCZ}-based \ac{OFDM} scheme for short-packet transmission implemented with frequency-mapping and both the \ac{ML} and \ac{dizet} decoders. We simulate the transmission of \ac{OFDM} symbols through both an \ac{AWGN} and fading channel. In the \ac{AWGN} channel, signals are perturbed by a noise vector $\bs{\mathrm{w}}$ with elements drawn from a complex normal distribution having zero-mean and variance $N_0$, i.e., $w_i\sim\mathcal{CN}(0,N_0)$. In the fading channel, subcarriers of the $i$th OFDM symbol are multiplied by the same single-tap complex gain $H_i\sim\mathcal{CN}(0,1)$ before being perturbed by \ac{AWGN}.\footnote{The authors note that because each subcarrier is multiplied by the same complex gain, $L=L_\mathrm{eff}=1$, and the performance of time- and frequency-mapping are rendered equivalent.} For each $K$, the \ac{IDFT} size is $\Nidft=2^K$, and the radius used to generate codebooks is computed according to (\ref{eq:dizetR}) and (\ref{eq:mlR}) for the \ac{dizet} and \ac{ML} decoders, respectively.
	
	Fig.~\ref{fig:awgnErrorCurves} shows the \ac{BER} and \ac{BLER} curves simulated in the \ac{AWGN} channel for \ac{BMOCZ} with $K\in\{4,7,10\}$. The \ac{ML} decoder outperforms the \ac{dizet} decoder in terms of both \ac{BER} and \ac{BLER} by a margin greater than 5 dB. This result is expected as the polynomial coefficients of \ac{BMOCZ} are more robust against additive noise than the zeros \cite{walk2019principles,wilkinson1984perfidious}. Fig.~\ref{fig:fadingErrorCurves} pictures the \ac{BER} and \ac{BLER} curves simulated in the fading channel for \ac{BMOCZ} with $K\in\{4,7,10\}$. While the \ac{BER} performance of the two decoders is comparable, notice that the \ac{ML} decoder again outperforms the \ac{dizet} decoder in \ac{BLER}, this time by a margin of roughly 3~dB. Thus, for suitable choices of the radius $R$, the \ac{ML} decoder achieves more reliable communication than the \ac{dizet} decoder for \ac{BMOCZ}.
	
	\subsection{Error Rates for Different Subcarrier Mappings} \label{subsec:mapping_performance}
	
	This section compares the performance of the \ac{OFDM} subcarrier mappings in a fading channel. We consider a decaying exponential power-delay profile (PDP) having $L$ taps given by 
	\begin{equation}
		\rho_l=\mathbb{E}\left[|h_l|^2\right]=\frac{1-\rho}{1-\rho^L}\rho^l,
	\end{equation}
	where $l\in\{0,1,\dots,L-1\}$ and $\rho\in(0,1)$ is the profile's decay constant \cite{walk2019principles}. Simulations are performed with packets comprising $P=64$ polynomial sequences, $L=K=9$, $\Nidft=256$, and $\rho=0.2$. The number of \ac{OFDM} symbols and subcarriers considered for time-frequency mapping are $M=5$ and $\mathcal{L}=128$, respectively. For the \ac{dizet} and \ac{ML} decoders, we again generate \ac{BMOCZ} codebooks using the corresponding radii in (\ref{eq:dizetR}) and (\ref{eq:mlR}).
	
\begin{figure}[t]
		\centering
		\subfloat[Bit error rate performance in a fading channel.]{\includegraphics[width=3in]{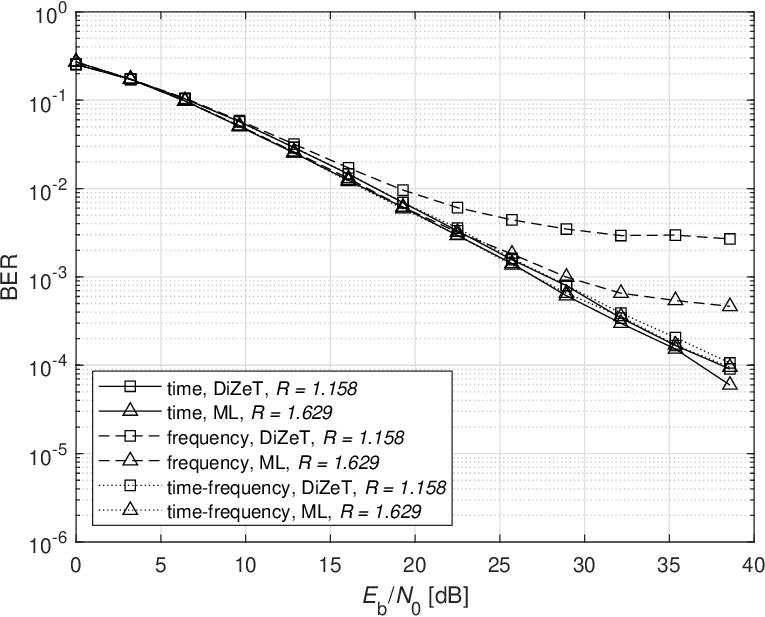}\label{subfig:ofdmBER}}~\\
		\subfloat[Block error rate performance in a fading channel.]{\includegraphics[width=3in]{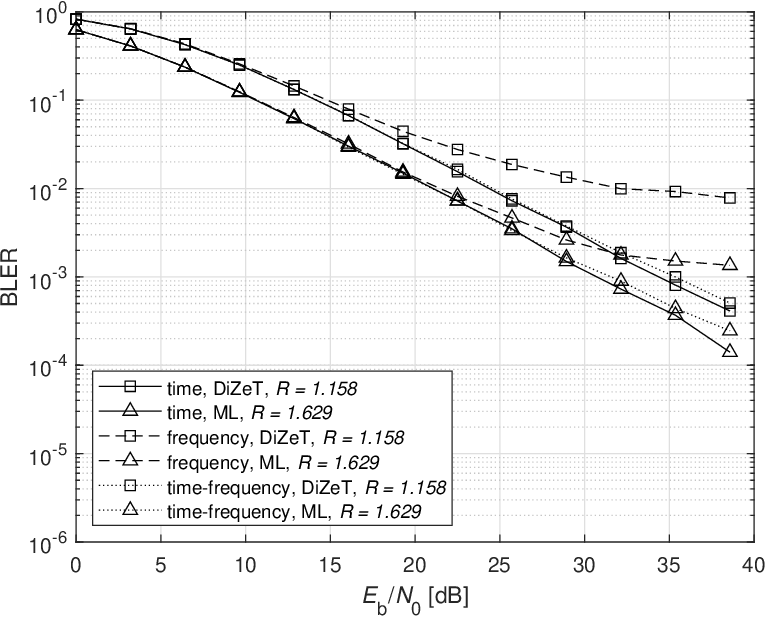}\label{subfig:ofdmBLER}}	
		\caption{Comparison of the \ac{OFDM} subcarrier mappings in a fading channel for $L=K=9$ and $\rho=0.2$. The radius $R$ is chosen to maximize codeword and zero separation for the \ac{ML} and \ac{dizet} decoders, respectively.}	
		\label{fig:ofdmComparison}
	\end{figure}

	Fig.~\ref{fig:ofdmComparison} shows the \ac{BER} and \ac{BLER} curves of the three \ac{OFDM} waveforms simulated in the fading channel. As anticipated from the discussions in \ref{subsec:mappings}, the time-mapping waveform outperforms the frequency-mapping waveform as each polynomial sequence observes flat-fading. In fact, the waveform with frequency-mapping experiences an error floor due to distortion of the polynomial coefficients induced by the frequency-selective channel. Similar performance is observed between time-frequency mapping and time-mapping. Given its heightened flexibility, however, we regard the time-frequency waveform as the most practical for real implementations, provided that \ac{PAPR} is not a primary design constraint.
	
	\subsection{\ac{PAPR} for Different Subcarrier Mappings} \label{subsec:papr}
	
\begin{figure}[t]
		\centering
		\includegraphics[width=3in]{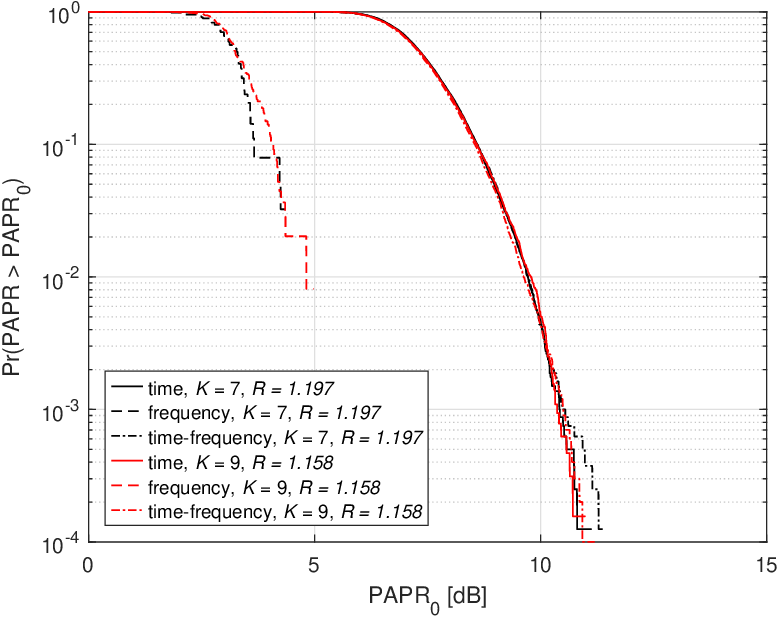}
		\caption{\ac{CCDF} of \ac{PAPR} for \ac{BMOCZ}-based \ac{OFDM} waveforms employing different sequence-to-subcarrier mappings.}	
		\label{fig:papr}
	\end{figure}

	In this section, we compare the \ac{PAPR} of the described subcarrier mappings for \ac{BMOCZ}-based \ac{OFDM}. The \ac{PAPR} of a discrete-time signal $s[n]$ with length $N_\text{s}$ is computed as 
	\begin{equation}
		\PAPR\lp s[n]\rp_\mathrm{dB}=10\log_{10}\lp\frac{\max|s[n]|^2}{\Pavg}\rp,
	\end{equation}
	where $\Pavg$ is the average power of the signal, i.e.,
	\begin{equation}
		\Pavg = \frac{1}{N_\text{s}}\sum_{n=0}^{N_\text{s}-1}|s[n]|^2.
	\end{equation}
	For our simulations, a single packet consists of $P=64$ polynomial sequences. The \ac{IDFT} size for each mapping strategy is $\Nidft=512$, and a \ac{QPSK} scrambler is implemented for time-mapping and time-frequency mapping to reduce \ac{PAPR}. Fig.~\ref{fig:papr} shows the \ac{CCDF} of \ac{PAPR} computed for $K\in\{7,9\}$ and the radius $R$ from (\ref{eq:mlR}). While there appears no considerable difference between \ac{PAPR} for time-mapping and time-frequency mapping, \ac{PAPR} is indeed minimized by mapping the polynomial sequences directly to frequency.
	
	\section{Concluding Remarks} \label{sec:conclusion}
	
	In this work, we investigate the radius for \ac{BMOCZ} that maximizes communication reliability. We first show that different radii maximize zero and codeword separation and that the optimal radius for \ac{BMOCZ} is a function of the utilized decoder (i.e., \ac{ML} or \ac{dizet}). Simulations of a \ac{BMOCZ}-based \ac{OFDM} scheme in \ac{AWGN} and fading channels validate that the \ac{ML} decoder outperforms the \ac{dizet} decoder for suitable choices of the radius parameter. Finally, we discuss the merits of different sequence-to-subcarrier mappings for \ac{BMOCZ}-based \ac{OFDM} to limit the number of zeros introduced by the channel. We propose a flexible time-frequency \ac{OFDM} waveform that is robust against a frequency-selective channel at the expense of increased \ac{PAPR}. Future work will focus on the optimal radius for \ac{MOCZ} more generally, the derivation of an expression for the radius maximizing codeword separation, and further quantitative analysis of the introduced \ac{BMOCZ}-based \ac{OFDM} waveforms.
	
	\bibliographystyle{IEEEtran}
	\bibliography{references}

\begin{thebibliography}{10}
\providecommand{\url}[1]{#1}
\csname url@samestyle\endcsname
\providecommand{\newblock}{\relax}
\providecommand{\bibinfo}[2]{#2}
\providecommand{\BIBentrySTDinterwordspacing}{\spaceskip=0pt\relax}
\providecommand{\BIBentryALTinterwordstretchfactor}{4}
\providecommand{\BIBentryALTinterwordspacing}{\spaceskip=\fontdimen2\font plus
\BIBentryALTinterwordstretchfactor\fontdimen3\font minus
  \fontdimen4\font\relax}
\providecommand{\BIBforeignlanguage}[2]{{%
\expandafter\ifx\csname l@#1\endcsname\relax
\typeout{** WARNING: IEEEtran.bst: No hyphenation pattern has been}%
\typeout{** loaded for the language `#1'. Using the pattern for}%
\typeout{** the default language instead.}%
\else
\language=\csname l@#1\endcsname
\fi
#2}}
\providecommand{\BIBdecl}{\relax}
\BIBdecl

\bibitem{bockelmann2016massive}
C.~Bockelmann, N.~Pratas, H.~Nikopour, K.~Au, T.~Svensson, C.~Stefanovic,
  P.~Popovski, and A.~Dekorsy, ``Massive machine-type communications in 5{G}:
  Physical and {MAC}-layer solutions,'' \emph{IEEE Communications Magazine},
  vol.~54, no.~9, pp. 59--65, 2016.

\bibitem{popovski2014ultra}
P.~Popovski, ``Ultra-reliable communication in 5{G} wireless systems,'' in
  \emph{Proc. 1st International Conference on 5{G} for Ubiquitous
  Connectivity}.\hskip 1em plus 0.5em minus 0.4em\relax IEEE, 2014, pp.
  146--151.

\bibitem{nawaz2021non}
S.~J. Nawaz, S.~K. Sharma, B.~Mansoor, M.~N. Patwary, and N.~M. Khan,
  ``Non-coherent and backscatter communications: Enabling ultra-massive
  connectivity in 6{G} wireless networks,'' \emph{IEEE Access}, vol.~9, pp.
  38\,144--38\,186, 2021.

\bibitem{witrisal2009noncoherent}
K.~Witrisal, G.~Leus, G.~J. Janssen, M.~Pausini, F.~Tr{\"o}sch, T.~Zasowski,
  and J.~Romme, ``Noncoherent ultra-wideband systems,'' \emph{IEEE Signal
  Processing Magazine}, vol.~26, no.~4, pp. 48--66, 2009.

\bibitem{xu2019sixty}
C.~Xu, N.~Ishikawa, R.~Rajashekar, S.~Sugiura, R.~G. Maunder, Z.~Wang, L.-L.
  Yang, and L.~Hanzo, ``Sixty years of coherent versus non-coherent tradeoffs
  and the road from 5{G} to wireless futures,'' \emph{IEEE Access}, vol.~7, pp.
  178\,246--178\,299, 2019.

\bibitem{walk2017short}
P.~Walk, P.~Jung, and B.~Hassibi, ``Short-message communication and {FIR}
  system identification using {H}uffman sequences,'' in \emph{Proc. IEEE
  International Symposium on Information Theory (ISIT)}.\hskip 1em plus 0.5em
  minus 0.4em\relax IEEE, 2017, pp. 968--972.

\bibitem{walk2019principles}
------, ``{MOCZ} for blind short-packet communication: Basic principles,''
  \emph{IEEE Transactions on Wireless Communications}, vol.~18, no.~11, pp.
  5080--5097, 2019.

\bibitem{walk2020practical}
P.~Walk, P.~Jung, B.~Hassibi, and H.~Jafarkhani, ``{MOCZ} for blind
  short-packet communication: Practical aspects,'' \emph{IEEE Transactions on
  Wireless Communications}, vol.~19, no.~10, pp. 6675--6692, 2020.

\bibitem{ackroyd1970design}
M.~H. Ackroyd, ``The design of {H}uffman sequences,'' \emph{IEEE Transactions
  on Aerospace and Electronic Systems}, no.~6, pp. 790--796, 1970.

\bibitem{siddiqui2023spectrally}
A.~A. Siddiqui, E.~Bedeer, H.~H. Nguyen, and R.~Barton, ``Spectrally-efficient
  modulation on conjugate-reciprocal zeros ({SE-MOCZ}) for non-coherent short
  packet communications,'' \emph{IEEE Transactions on Wireless Communications},
  2023.

\bibitem{sasidharan2024alternative}
B.~Sasidharan, E.~Viterbo, and Y.~Hong, ``Alternative zero codebooks for {MOCZ}
  with reduced {PAPR},'' \emph{IEEE Communications Letters}, 2024.

\bibitem{sun2023noncoherent}
Y.~Sun, Y.~Zhang, G.~Dou, Y.~Lu, and Y.~Song, ``Noncoherent {SIMO} transmission
  via {MOCZ} for short packet-based machine-type communications in
  frequency-selective fading environments,'' \emph{IEEE Open Journal of the
  Communications Society}, 2023.

\bibitem{walk2021multi}
P.~Walk and W.~Xiao, ``Multi-user {MOCZ} for mobile machine type
  communications,'' in \emph{Proc. IEEE Wireless Communications and Networking
  Conference (WCNC)}.\hskip 1em plus 0.5em minus 0.4em\relax IEEE, 2021, pp.
  1--6.

\bibitem{forney1972maximum}
G.~Forney, ``Maximum-likelihood sequence estimation of digital sequences in the
  presence of intersymbol interference,'' \emph{IEEE Transactions on
  Information Theory}, vol.~18, no.~3, pp. 363--378, 1972.

\bibitem{wilkinson1984perfidious}
J.~H. Wilkinson, ``The perfidious polynomial,'' \emph{Studies in Numerical
  Analysis}, vol.~24, pp. 1--28, 1984.

\end{thebibliography}
	
\end{document}